\documentclass[10pt, journal,twoside]{IEEEtran}

\usepackage{graphicx}
\usepackage{subfigure}
\usepackage{amsmath}
\usepackage{amssymb}
\usepackage{cases}
\usepackage{times} 
\usepackage[font=footnotesize,labelfont=bf]{caption}
\usepackage{caption}
\usepackage{algorithm}
\usepackage{algorithmicx}
\usepackage{algpseudocode}
\usepackage{color}
\usepackage{multirow}
\usepackage{flushend}

\makeatletter
  \newcommand\figcaption{\def\@captype{figure}\caption}
  \newcommand\tabcaption{\def\@captype{table}\caption}
\makeatother
\usepackage{array} 
\newcolumntype{C}[1]{>{\centering}p{#1}}
\begin{document}
 \title{EdgeLeakage: Membership Information Leakage in Distributed Edge Intelligence Systems}
 
 \author{Kongyang Chen, Yi Lin, Hui Luo, Bing Mi, Yatie Xiao, Chao Ma, and Jorge S\'{a} Silva
\IEEEcompsocitemizethanks{
\IEEEcompsocthanksitem K. Chen, Y. Lin, H. Luo, and Y. Xiao are with Guangzhou University, Guangzhou, China.  K. Chen is also with Pazhou Lab, Guangzhou, China.
\IEEEcompsocthanksitem B. Mi is with Guangdong University of Finance and Economics, and also with Pazhou Lab, Guangzhou, China.
\IEEEcompsocthanksitem C. Ma is with Wuhan University, Wuhan, China.
\IEEEcompsocthanksitem J. S\'{a} Silva is with University of Coimbra, INESC Coimbra, Portugal.
}}

\IEEEtitleabstractindextext{
\begin{abstract}
In contemporary edge computing systems, decentralized edge nodes aggregate unprocessed data and facilitate data analytics to uphold low transmission latency and real-time data processing capabilities. Recently, these edge nodes have evolved to facilitate the implementation of distributed machine learning models, utilizing their computational resources to enable intelligent decision-making, thereby giving rise to an emerging domain referred to as edge intelligence. However, within the realm of edge intelligence, susceptibility to numerous security and privacy threats against machine learning models becomes evident. This paper addresses the issue of membership inference leakage in distributed edge intelligence systems. Specifically, our focus is on an autonomous scenario wherein edge nodes collaboratively generate a global model. The utilization of membership inference attacks serves to elucidate the potential data leakage in this particular context. Furthermore, we delve into the examination of several defense mechanisms aimed at mitigating the aforementioned data leakage problem. Experimental results affirm that our approach is effective in detecting data leakage within edge intelligence systems, and the implementation of our defense methods proves instrumental in alleviating this security threat. Consequently, our findings contribute to safeguarding data privacy in the context of edge intelligence systems.
\end{abstract}
}
\maketitle
\IEEEdisplaynontitleabstractindextext
\IEEEpeerreviewmaketitle

\section{Introduction}
In modern edge computing systems, distributed edge nodes collect raw information and provide data analytics to support low transmission latency and real-time data processing. Recently, edge nodes can provide distributed machine learning models with their available computation resources to support an intelligent decision making, inspiring an emerging area called edge intelligence.
Traditional machine learning depends on a large amount of data samples to support its training, and it usually needs a central server for data collection, model training or aggregation, etc. Considering the critical privacy concerns, many organizations are not allowed to share their individual data, which thus significantly decreases the overall model accuracy.  Therefore, Federated Learning (FL) is proposed to server as a novel distributed learning paradigm, where each participant client user keeps its own individual data locally, and only shares its model parameters (or gradient updates) to a centralized server for model aggregation~\cite{DBLP:conf/aistats/McMahanMRHA17}. However, existing solutions show that Federated learning still suffers from serious information leakage when the centralized server is attacked. 
With a remote central server, it is thus hard to build a connection-frequency model training such as Federated learning. To improve the privacy and security, each client has a chance to be chosen as a temporary server to aggregate model updates from participant clients. 

In the realm of model security, it has been established that machine learning models face vulnerability to various model attacks, such as membership inference attacks~\cite{DBLP:conf/icml/YurochkinAGGHK19, DBLP:journals/corr/abs-1708-06145, DBLP:conf/sp/NasrSH19}, model inversion attacks~\cite{DBLP:conf/ccs/YangZCL19}, and property inference attacks~\cite{DBLP:conf/ccs/GanjuWYGB18}. These attacks have the potential to result in the leakage of sensitive information from the training dataset. For instance, a membership inference attack determines whether a given data sample was utilized in the previous model training process. Such knowledge is advantageous for adversaries seeking to exploit model security, posing potential severe ramifications, especially in the deployment of Machine Learning as a Service (MLaaS)~\cite{DBLP:conf/icml/YurochkinAGGHK19}. Concerning the membership inference attack (MIA), Shokri et al.\cite{DBLP:journals/corr/ShokriSS16} introduced this attack against machine learning models employing shadow models in black-box scenarios, effectively transforming MIA into a binary-classification problem. Additionally, efforts have been made to address the cost associated with such attacks. Specifically, research by Yeom et al.\cite{DBLP:journals/corr/abs-1806-01246} and Song and Mittal~\cite{DBLP:conf/uss/SongM21} delved into metric-based membership inference attacks, focusing on factors such as prediction confidence and prediction entropy, respectively, aiming to mitigate the attack's computational expense.

In this study, we investigate security threats within distributed edge intelligence systems. We specifically focus on employing membership inference attacks to elucidate potential data leakage, encompassing NN-based attacks, Metric-based attacks, and Differential attacks. Furthermore, we assess the performance of these attacks on diverse participant client users, with experimental results substantiating the existence of potential data leakage. Finally, we introduce several defense mechanisms aimed at preempting the aforementioned attacks. The principal contributions of our research are delineated as follows:

\begin{itemize}
	\item We analyze the security model within distributed edge intelligence systems and demonstrate the integration of various membership inference attack methods, including NN-based attacks, Metric-based attacks, and Differential attacks. Additionally, we propose several defense strategies to counteract these attacks.
	\item Experimental findings validate the efficacy of our approach in detecting data leakage issues within edge intelligence systems, while also highlighting the utility of our defense mechanisms in mitigating this security threat.
\end{itemize}

The subsequent sections of this paper are structured as follows: Section~\ref{Overview} provides an overview of the framework underpinning our membership inference attacks. Section~\ref{Experiments} presents the results obtained from our experiments. Section~\ref{Defense} outlines several defense methodologies devised to combat these attacks. Section~\ref{Related} delves into related research, while Section~\ref{Conclusion} furnishes a comprehensive conclusion to our study.

\section{Membership Inference Leakage against Distributed Edge Intelligence Systems}\label{Overview}
In this section, we present the system framework of our distributed edge intelligence system and propose our membership inference attack model against it.

\subsection{Distributed Edge Intelligence Systems}
\begin{figure}[!t]
	\centering
	\includegraphics[width=0.45\textwidth]{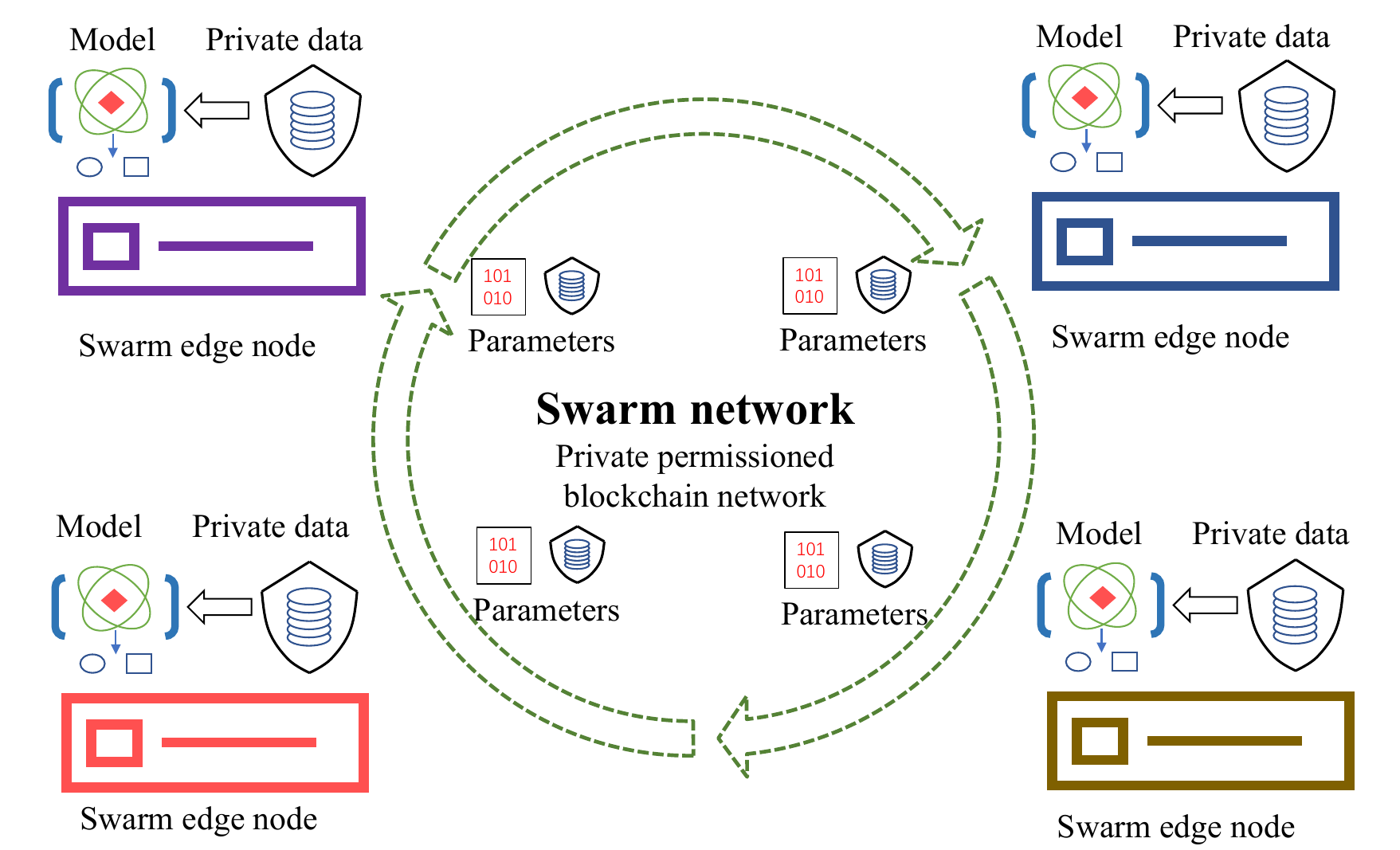}
	\caption{Distributed Edge Intelligence Systems.} \label{fig1}
\end{figure}

As depicted in Figure~\ref{fig1}, our distributed edge intelligence system distributes data samples across edge clients, with each client sharing and aggregating local model updates for joint training. Unlike centralized data aggregation methods such as federated learning, our approach does not necessitate a specific central server for parameter sharing and aggregation. Instead, each client user can function as a central client, selected before each aggregation through an internally determined server selection mechanism. This strategy mitigates various potential attacks and failures, including central client failures or privacy breaches. Additionally, our system leverages edge computing and blockchain technology to bolster transmission security. Given that edge nodes learn in a distributed manner without central server assistance, we refer to this approach as "swarm learning" for brevity.
\subsection{MIA against Distributed Edge Intelligence Systems}
In the membership inference attack (MIA) against our system, we consider the attacker to be one of the internal clients participating in our distributed edge system, with the objective of targeting other clients within the same system. It is assumed, without loss of generality, that the attacker has access to certain information about the distributed edge system, such as the model architecture. Consequently, both white-box and black-box attacks are viable within this context. We employ the membership inference attack technique described in~\cite{DBLP:journals/corr/abs-1806-01246}, which utilizes a single shadow for conducting the attack.

\begin{figure*}[!t]
	\centering
	\includegraphics[width=0.85\textwidth]{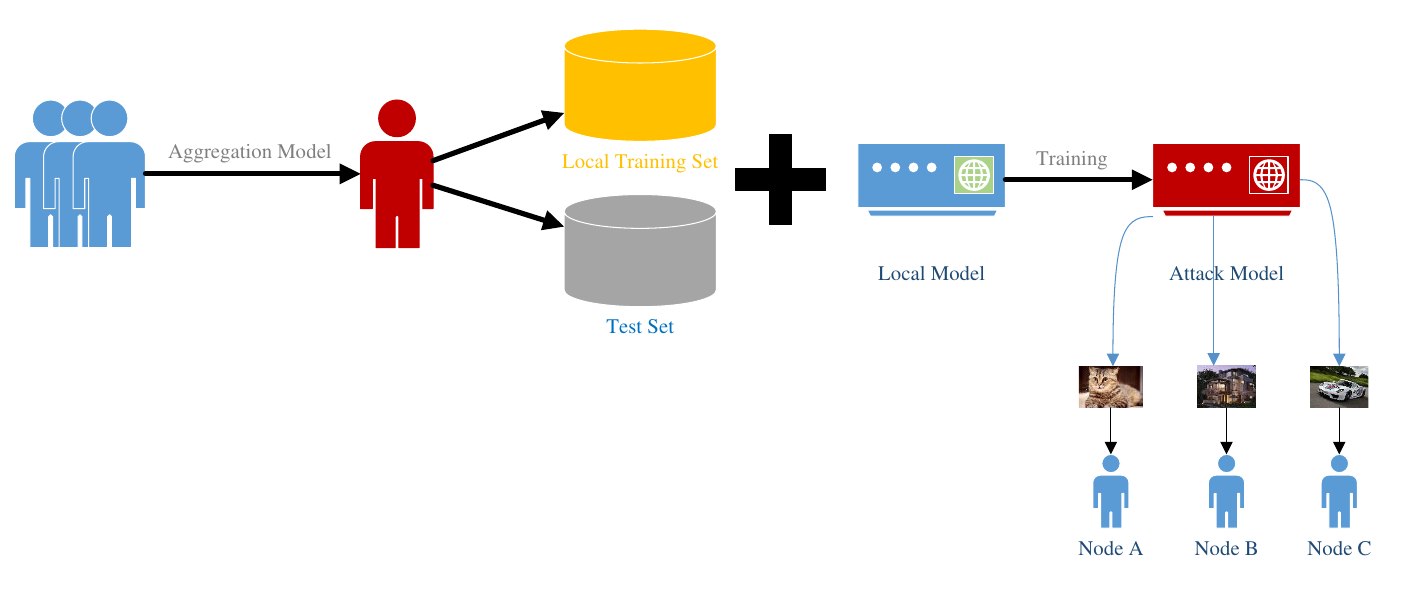}
	\caption{Our Attack Architecture.} \label{fig2}
\end{figure*}

For the attacking client, the shadow training set comprises its local training and test sets, while the shadow model corresponds to its local model. Thus, there is no necessity to train any additional model aside from the attack model. The architecture of our attack is illustrated in Figure~\ref{fig2}. It is important to note that while the attacking client participates in the global model aggregation as a regular client, it also endeavors to gather local data information from other clients in a malicious manner.

We will introduce three membership inference attack against our distributed edge system In this following. 

\subsection{Attack 1: NN-based Attack}\label{NN-based Attack}
\begin{figure*}[!t]
	\centering
	\includegraphics[width=0.85\textwidth]{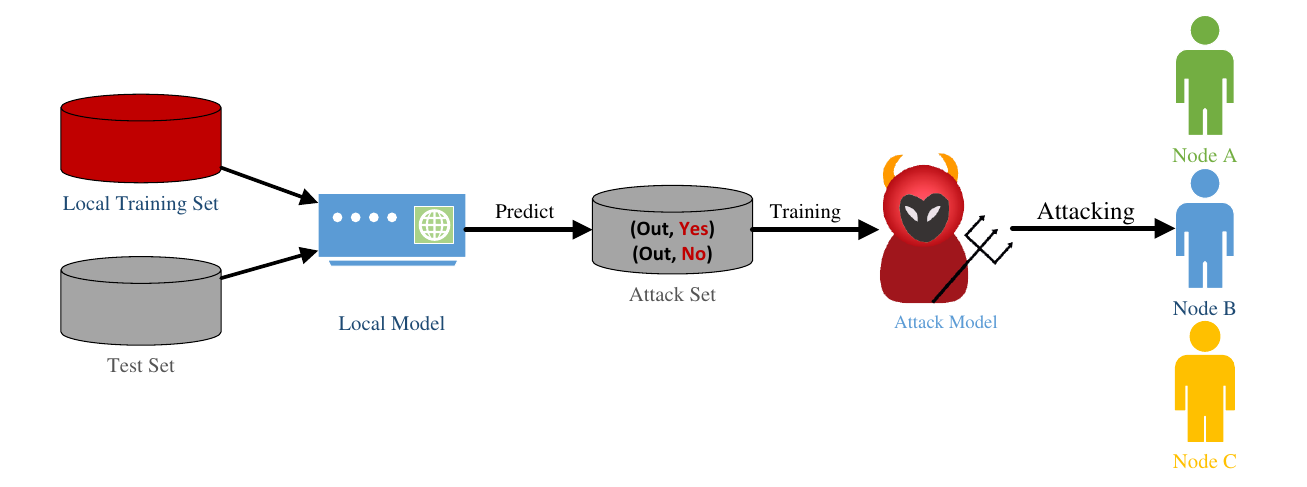}
	\caption{Improved Shadow Model Attack Architecture.} \label{fig}
\end{figure*}

In this section, we train an attack model at the attacking client to target other clients. Illustrated in Figure~\ref{fig}, our refined shadow model attack architecture is depicted. To elucidate our attack methodology, we assume the presence of $N$ clients in the distributed edge system, where only the last client (i.e., with client ID $N$) acts as the malicious attacking client. Thus, the objective is for the last client to target the first client, represented simply as \textit{$N$ attack 1}, or \textit{$N \rightarrow$ 1}. Furthermore, we assess the transmission of the attack across these $N$ clients, with client $i$ acting as the malicious attacker, where $i$ ranges from 2 to $N-1$. These attack scenarios can be denoted as \textit{$i$ attack 1}, or \textit{$i \rightarrow$ 1}, where $i$ ranges from 2 to $N-1$. Additionally, we utilize both balanced and unbalanced datasets to evaluate the attack performance of the attack model.

\subsection{Attack 2: Metric-based Attack}
In this experiment, we implement two metric-based attacks on our distributed edge system: attacks based on prediction confidence and attacks based on prediction entropy. Unlike the NN-based attacks discussed in Section~\ref{NN-based Attack}, these attacks rely on metrics that incur lower costs and overhead. As their names suggest, we train an attack model and observe the differences in predictions made by the model for various data samples. Specifically, we conduct membership inference attacks on our distributed edge system with 2, 3, and 4 clients.

We assess the effectiveness of metric-based attacks using the CIFAR-10 dataset, dividing it into  $D^{Train}_{n,SL}$ and $D^{Test}_{SL}$. Here, $D^{Train}{n,SL}$ represents the training sets of client $n$, while all clients utilize the same test set $D^{Test}_{SL}$. Given that each client within the same edge framework employs identical model architectures, we utilize the model of the attacking client itself as the basis for measuring attack-based metrics. Consequently, we designate  $D^{Train}_{n,SL}$  as positive samples (i.e., member labels) in the training sets for the attack model, and $D^{Test}_{SL}$ as negative samples (i.e., non-member labels).

For this experiment, we employ ResNet50, a widely used model in image recognition, as the training model for our distributed edge clients. We conduct membership inference attacks based on prediction entropy and prediction confidence on our edge frameworks with 2, 3, and 4 clients. All experiments in this section are categorized into two classes for the attack model: distinguishing only between members and non-members.

\subsection{Attack 3: Differential Attack}
In the section, we introduce Maximum Mean Discrepancy (MMD), which is crucial for differential attacks. Furthermore, we present the dataset utilized in our attack.

The core concept behind differential attacks lies in Maximum Mean Discrepancy (MMD). MMD, introduced by Gretton et al. \cite{DBLP:journals/jmlr/GrettonBRSS12}, is a smooth function that tests whether distributions \( p \) and \( q \) differ. A large value of this function suggests potential differences between the distributions. Mathematically, it is defined as:

\[ F(D^{Mem}_{SL}, D^{Nonmem}_{SL}) = ||\frac{1}{n}\sum_{i=1}^{n_t}\phi(y_i)-\frac{1}{n_s}\sum_{j=1}^{n_s}\phi(y^{'}_j)||_v \]

\noindent where \( y_i \in D^{Mem}_{SL} \), \( y^{'}_j \in D^{Nonmem}_{SL} \), \( n_t \) and \( n_s \) represent the sizes of \( D^{Mem}_{SL} \) and \( D^{Nonmem}_{SL} \) respectively, \( v \) denotes the dimension of the kernel space, and \( \phi \) is a feature space map defined as \( k \rightarrow v \). In our experiment, we employ the Gaussian kernel function \( k(y, y') = \langle\phi(y), \phi(y')\rangle = exp(-\frac{||y-y'||}{2\sigma^2}) \).

With the proliferation of defense methods such as Adversarial Regularization \cite{DBLP:conf/ccs/NasrSH18}, MemGuard \cite{DBLP:conf/ccs/JiaSBZG19}, and differential privacy \cite{DBLP:journals/corr/AbadiCGMMTZ16}, effectively acting on models, it becomes challenging to directly differentiate between members and non-members through the model prediction probability space. Hence, we map it to the Reproducing Kernel Hilbert Space (RKHS) \cite{DBLP:conf/ismb/BorgwardtGRKSS06} and then calculate the distance between two centroids in the kernel space.

Differential comparison, an idea applied to machine learning in recent years \cite{DBLP:journals/corr/abs-2101-01341}, has been introduced to membership inference attacks. In this experiment, we propose an enhanced differential attack against our distributed edge system. For the training datasets of distributed edge clients, we utilize both independently and non-independently identically distributed cases respectively to evaluate the impact of distribution variations on distributed edge privacy leaks.

The differential attack employs Maximum Mean Discrepancy to measure the characteristic distance between two groups of samples: one group closely resembling the training sample, and the other closely resembling the non-training sample. In the first variant of the differential attack, denoted as differential attack one, we iteratively add the target sample to \( D^{Mem}_{SL} \) and determine the target sample's membership by comparing the distance between \( D^{Mem}_{SL} \) and \( D^{Nonmem}_{SL} \). Details of the differential attack are shown in Algorithm~\ref{alg:attack}. Lines 1-5 of Algorithm~\ref{alg:attack} initialize some variables, while lines 6-14 represent the iterative calculation process for determining the target sample's membership. Subsequent lines detail the discrimination methods employed in the attack.
\begin{algorithm}[!t]
	\caption{ Differential attack against our system.}
	\label{alg:attack}
	\begin{algorithmic}[1]
		\Require
		$D^{Train}_{n,SL}$, $D^{Test}_{SL}$, $D^{Pre}_{Diff,SL}$
		\Ensure
		$T^{Mem}_{Pre}$, $T^{Nonmem}_{Pre}$
		\State $D^{Pre}_{Diff,SL}$←empty 
		\State flag←n
		\While{flag}
		\State state←0
		\State statev←0
		\For{$D^{Mem}_{SL}\in D^{Train}_{flag,SL}$, $D^{Nonmem}_{SL}\in D^{Test}_{SL}$}
		\State $d'$←$F(D^{Mem}_{SL}, D^{Nonmem}_{SL})$
		\If{$d'$ > statev}
		\State state←i
		\State statev←$d'$
		\EndIf
		\EndFor
		\State flag←flag-1
		\EndWhile
		\If{state>0}
		\State $T^{Mem}_{Pre}$←($D^{Pre}_{Diff,SL}$, state)
		\Else
		\State $T^{Nonmem}_{Pre}$←$D^{Pre}_{Diff,SL}$
		\EndIf
	\end{algorithmic}
\end{algorithm}

\begin{algorithm}[!t]
	\caption{ Differential attack against our system.}
	\label{alg:Attack2}
	\begin{algorithmic}[1]
		\Require
		$D^{Train}_{n,SL}$, $D^{Test}_{SL}$, $D^{Pre}_{Diff,SL}$
		\Ensure
		$T^{Mem}_{Pre,i}$, $T^{Nonmem}_{Pre}$
		\State $D^{Pre}_{Diff,SL}$←empty 
		\State flag←n
		\While{flag}
		\State state←0
		\State statev←0
		\For{$D^{Pre}_{SL1}\in D^{Train}_{state,SL}$, $D^{Pre}_{SL2}\in D^{Train}_{flag,SL}$}
		\State $d'$←$F(D^{Mem}_{SL}, D^{Nonmem}_{SL})$
		\If{$d'$>statev}
		\State state←i
		\State statev←$d'$
		\EndIf
		\EndFor
		\State flag←flag-1
		\EndWhile
		\If{statev>$F(D^{Pre}_{state,SL}, D^{Test}_{SL})$}
		\State $T^{Mem}_{Pre, state}$←$D^{Pre}_{Diff,SL}$
		\Else
		\State $T^{Nonmem}_{Pre}$←$D^{Pre}_{Diff,SL}$
		\EndIf
	\end{algorithmic}
\end{algorithm}

In the second differential attack, we employ a different strategy. We iteratively calculate the membership of the target sample to differential groups (each group is closer to \( D^{Train}_{i,SL} \)). Algorithm~\ref{alg:Attack2} details the procedure. Lines 1-5 prepare initial variables, lines 6-14 outline the iterative calculation process for determining the target sample's membership, and subsequent lines specify the discrimination methods employed in the attack. The key distinction between this algorithm and the previous one (Algorithm~\ref{alg:Attack2}) lies in the direct calculation of the membership of \( D^{Train}_{i,SL} \) to \( D^{Train}_{k,SL} \) (where \( i \) is not equal to \( k \)).

Maximum Mean Discrepancy (MMD) serves as one of the bases for differential attacks, increasing with distributional differences. Intuitively, the distribution of local datasets among distributed edge clients within the same architecture is likely to vary. For instance, in medical applications of distributed edge systems, clients may span the globe, leading to disparate data distributions between locations like Guangzhou and New York, potentially driven by ethnic disparities or differences in lifestyle. Consequently, assessing privacy risks of our distributed edge system under non-IID conditions may better reflect reality. Lastly, we experiment with altering the aggregation weight of each client to evaluate its impact on the attack results.

\section{Experiment Evaluation}\label{Experiments}
In this section, we implement the aforementioned membership inference attack against our distributed edge intelligence system and evaluate its performance.

\subsection{Experiment Datasets}
In our experiments, we utilize several common datasets to assess the privacy risks associated with MIA, including CIFAR-10, CIFAR-100, and News. We adhere to the preprocessing methods outlined in \cite{DBLP:journals/corr/abs-1806-01246} and \cite{De2021}.

\textbf{CIFAR-10:} CIFAR-10 is a widely used dataset in the field of image recognition, comprising images with a resolution of 32×32 pixels. It consists of 50,000 training images and 10,000 test images categorized into 10 classes. Notably, each class contains 5,000 images, ensuring even distribution across the dataset.

\textbf{CIFAR-100:} Similar to CIFAR-10, CIFAR-100 serves as a benchmark dataset for image recognition tasks. It comprises 50,000 images classified into 100 classes. Each class contains 500 training images and 100 test images, maintaining an even distribution across the dataset.

\textbf{News:} The News dataset is an internationally recognized standard dataset commonly used in classification, data mining, and information retrieval research. It consists of approximately 20,000 newsgroup documents categorized into 20 distinct newsgroups covering various topics.

\subsection{Experiment Settings}
We assess the privacy risks of the distributed edge system on CIFAR-10, CIFAR-100, and News datasets. For each dataset, we divide it into two parts: \( D_{SL}^{Train} \) (comprising \( D_{1,SL}^{Train} \), ..., \( D_{n,SL}^{Train} \)) and \( D_{MIA}^{Train} \). According to the attack strategy, \( D_{MIA}^{Train} \) is further subdivided into \( D_{Shadow}^{Train} \) and \( D_{Shadow}^{Test} \). \( D_{Shadow}^{Train} \), categorized into members and non-members based on whether they participate in the training of the attacked client, is utilized to train the attack model. For image datasets, we employ convolutional neural networks (CNNs) to construct the client's model \cite{DBLP:journals/corr/abs-1806-01246}. For text datasets, linear neural networks (NNs) are used \cite{DBLP:journals/corr/abs-1806-01246}.

We consider three typical membership inference attacks in our system. The first is the \textit{One-to-One attack}. In a distributed edge system with \( N \) clients, this implies that the client with client ID \( N \) is the malicious attacking client. For instance, \textit{\(N \rightarrow 1\)} signifies that client \( N \) will attack client 1. The second attack is the \textit{Multi-to-One attack}. In a distributed edge system with \( N \) clients, this denotes that the client with client ID \( i \) is the malicious attacking client, where \( i=2,3,\cdots,N-1 \). For example, \textit{\(i \rightarrow 1\)} indicates that client \( i \) will attack client 1. The third attack is the \textit{One-to-Multi attack}. In a distributed edge scenario with \( N \) clients, this indicates that the client with client ID \( i \), where \( i=2,3,\cdots,N-1 \), is the malicious attacking client and will target other clients.

\begin{figure}[!t]
	\centering
	\subfigure[a][CIFAR-10 dataset.]{\includegraphics[width=0.45\textwidth]{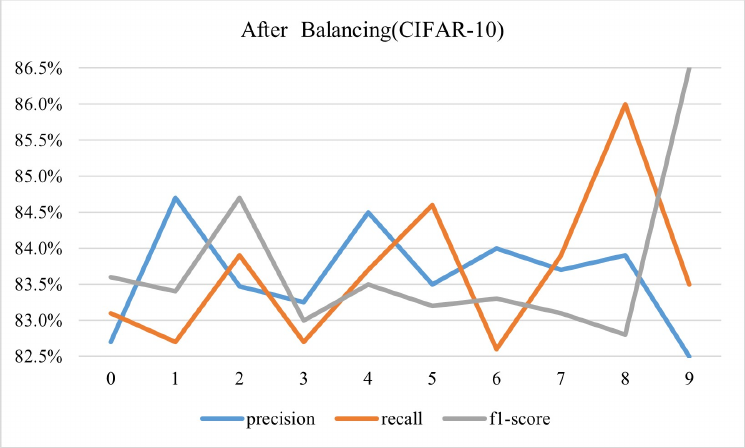}}\label{a} \\
	\subfigure[b][News dataset.]{\includegraphics[width=0.45\textwidth]{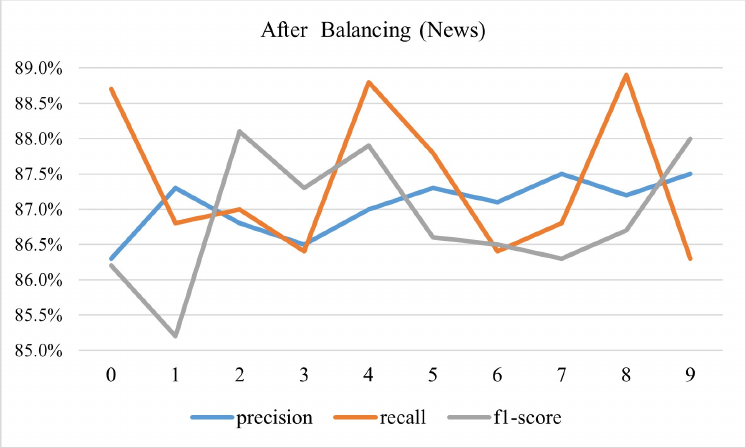}}\label{b}
	\caption{The attack results of \textit{One-to-One attack}.}
	\label{fig4}
\end{figure}

\begin{figure}[!t]
	\centering
	\includegraphics[width=0.45\textwidth]{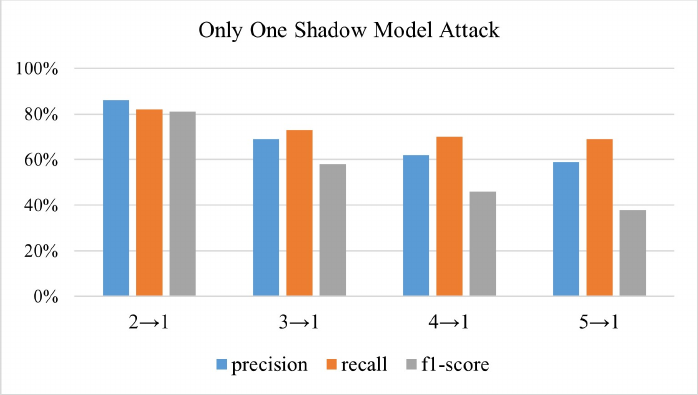}
	\caption{The attack results of \textit{One-to-One attack} on the CIFAR-10 dataset,  where \textit{$N \rightarrow$ 1} means that client $N$ attacks client 1.}
	 \label{fig3}
\end{figure}

\subsection{Experiment Results for Attack 1}
The attack performance of the \textit{One-to-One attack} is illustrated in Figure~\ref{fig3} and Figure~\ref{fig4}.
Figure~\ref{fig4} indicates that the attack results of the \textit{One-to-One attack} on the CIFAR-10 dataset and News dataset are consistently stable, achieving high attack performance (>82\%) across varying numbers of clients in our distributed edge system. However, Figure~\ref{fig3} suggests that the effectiveness of the attack diminishes as the size of the distributed edge system increases. For instance, on the CIFAR-10 dataset, the attack results decrease from 83\% to 40\% (even lower than the blind guess baseline) as the client size increases from 2 to 5.

\begin{figure}[!t]
	\centering
	\includegraphics[width=0.45\textwidth]{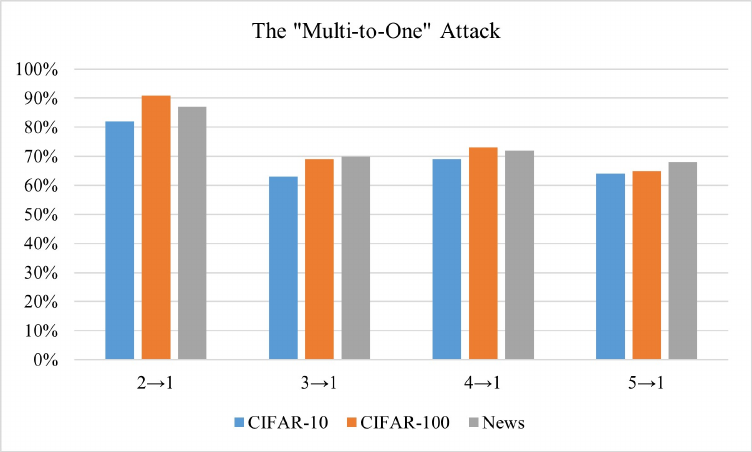}
	\caption{The attack results of \textit{Multi-to-One},  where the \textit{$i \rightarrow$ 1} means that client $i$ attacks client 1.}
	 \label{fig5}
\end{figure}

The attack performance of the \textit{Multi-to-One attack} is depicted in Figure~\ref{fig5}. It can be observed that each client in the distributed edge system can successfully execute the membership inference attack. However, there is a slight decrease in attack accuracy as the client size increases from 3 to 5.

\begin{figure}[!t]
	\centering
	\includegraphics[width=0.45\textwidth]{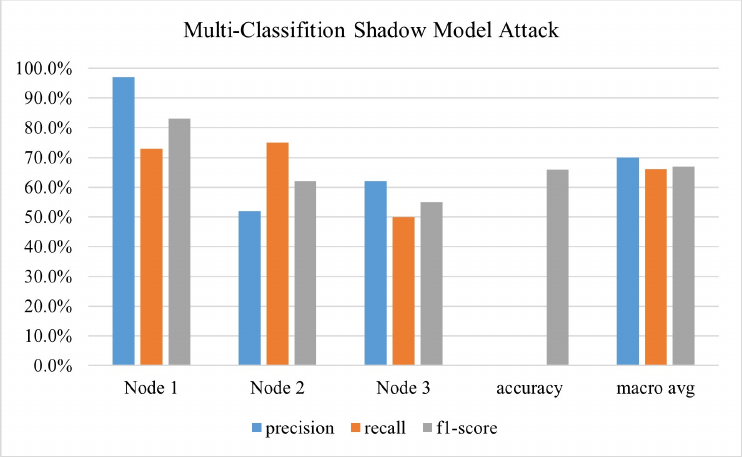}
	\caption{The attack results of \textit{One-to-Multi},  where \textit{accuracy} and \textit{macro-avg} are the general attack accuracy and macro-average result of the whole attack.}
	 \label{fig6}
\end{figure}

The attack performance of the \textit{One-to-Multi attack} is illustrated in Figure~\ref{fig6}. In our distributed edge system, we achieve an accuracy of 66\%, significantly surpassing the baseline of blind guessing (i.e., 33\% for a client size of \( N=3 \)).

We analyze the variations in \textit{One-to-One attack} results under balanced and unbalanced datasets and provide relevant explanations. On unbalanced datasets, the effectiveness of the attack decreases with an increase in the number of clients. This phenomenon seems reasonable because ordinary internal clients cannot observe the model aggregation process, preventing the attacking client from obtaining specific information about each client's parameter contributions. However, membership inference attacks often exploit the model's lack of generalization ability to infer membership based on performance disparities between training and non-training data. Without knowledge of the specific generalization differences of the attacked client, our attack effectiveness weakens with an increasing number of client points.

But why does the attack effectiveness improve after balancing the datasets? According to \cite{DBLP:conf/iclr/ZhangBHRV17}, a comprehensive study of deep learning models reveals that a "perfect" model can be achieved if the model parameters exceed the number of training and test sets. Additionally, they argue that overfitting and generalization abilities are not as intuitive as commonly perceived. Moreover, techniques like dropout and regularization are not always necessary and may not be as effective as using a simpler model directly. As per \cite{DBLP:conf/iclr/ZhangBHRV17}, our attack model contains far more parameters than the number of training and test sets, potentially leading to the learning of faults and redundant information. For instance, if there are more negative samples than positive samples in the training sets, the model may lean towards classifying low-confidence samples as negative. Furthermore, due to dataset errors and learning method issues, no "perfect" model exists in machine learning. This means that any model will misclassify data, and models with good classification performance may struggle with datasets containing misjudged data. Consequently, training and evaluating attacks on unbalanced datasets are scientifically unsound, potentially yielding results lower than the blind guess baseline (50\%). Balancing the positive and negative samples of the attack model leads to improvements in attack performance.

\subsection{Experiment Results for Attack 2}
\begin{figure}[!t]
	\centering
	\includegraphics[width=0.45\textwidth]{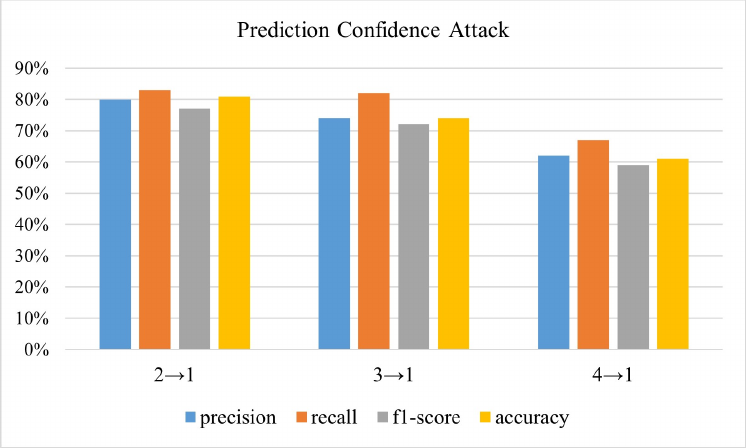}
	\caption{Prediction confidence attack.} \label{confidence}
\end{figure}

For the attack based on prediction entropy, we observe poor results on our distributed edge system. Even on the distributed edge system with only 3 clients, we achieve only 57\% accuracy, slightly better than the 50\% blind guess baseline, indicating weak performance. Figure~\ref{confidence} illustrates the results of the attack based on prediction confidence on our distributed edge system. Remarkably, even with 4 clients in the distributed edge system, we achieve a 66\% attack accuracy, significantly surpassing the 50\% blind guess baseline.

\begin{figure}[!t]
	\centering
	\includegraphics[width=0.45\textwidth]{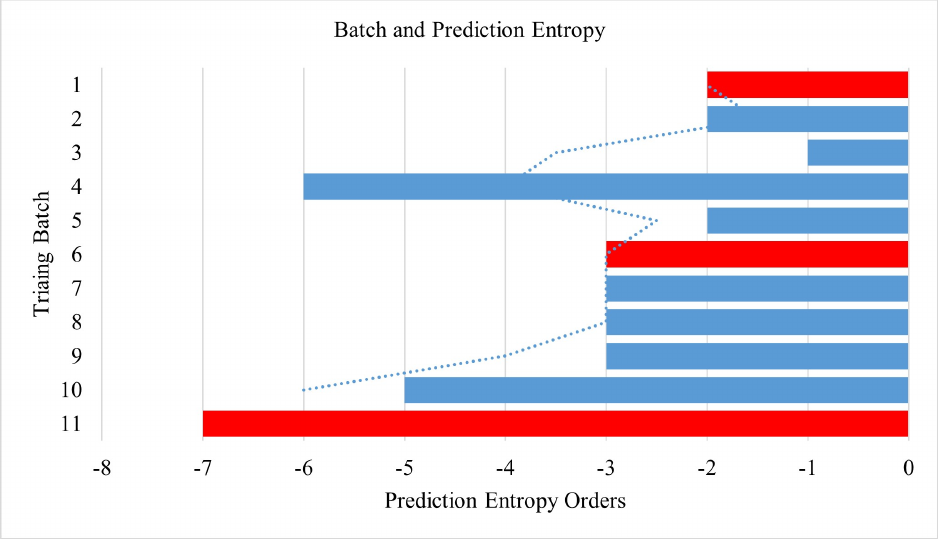}
	\caption{Relationship between training epochs and prediction entropy. The abscissa represents the magnitude of prediction entropy, while the ordinate denotes the training batch. The red batch indicates where the target sample is added.} \label{entropy}
\end{figure}

We attempt to elucidate and analyze the experimental results. Figure~\ref{entropy} illustrates the relationship between training epochs and prediction entropy. The red epochs indicate where target samples are added, while the blue epochs signify their removal. After each epoch of training, the post-training model is employed to predict these samples and obtain their prediction entropy.
The experimental findings reveal that the relationship between sample prediction entropy and training epochs is not consistently clear. While prediction entropy generally decreases with increasing training epochs, there are instances where the addition of target samples actually increases the prediction entropy of the model. Traditional membership inference attacks typically target conventional machine learning models that enhance prediction accuracy and generalization ability through numerous epochs. This aligns with the basis of attacks relying on prediction entropy.

However, our distributed edge system adopts a strategy of fewer epochs and more aggregation to enhance model prediction accuracy and generalization ability. Consequently, it exhibits a robust defense against membership inference attacks based on prediction entropy.
In contrast, attacks based on prediction confidence yield relatively better results. This may be attributed to the more complex relationship between model prediction performance and sample training status, compared to a simple entropy relationship. Although this relationship cannot be explicitly expressed currently, the attack model is able to capture it. Thus, attacks based on prediction confidence prove more effective against our distributed edge system.

\subsection{Experiment Results for Attack 3}
We utilize the CIFAR-100 dataset to evaluate and observe the performance of the differential attack against our distributed edge system. In the independent identically distributed (IID) experiment, we partition the dataset into two parts: \(D^{Train}_{n,SL}\) and \(D^{Test}_{SL}\), consistent with the settings in the previous experiment. For the non-IID experiment, we employ a method proposed by \cite{DBLP:conf/icml/YurochkinAGGHK19} to partition the dataset, generating non-IID datasets using Dirichlet distribution. Privacy evaluation is conducted on the distributed edge framework with 4 clients, where the datasets are divided into \(2n\) parts: \(D^{Train}_{n,SL}\) and \(D^{Test}_{SL}\). Each client's test set differs, and the size of the training set for each client is similar. Table~\ref{Non-IID} illustrates the distribution of the training set and test set for each client.

\begin{table*}[!t] 
	\centering
	\begin{tabular}{p{1cm}|p{8.5cm}|p{1cm}|p{1cm}}
		\hline 
		Client ID & Labels & Training Size & Test Size\\
		\hline
		1 & 4. 5. 6. 7. 9. 10. 14. 15. 16. 17. 19. 22. 31. 32. 
		34. 36. 37. 39. 42. 43. 46. 48. 50. 51. 54. 57. 58. 59. 
		60. 62. 63. 65. 66. 67. 68. 73. 74. 75. 76. 78. 79. 82. 
		83. 88. 90. 93. 94. 95. 97. 98. & 11360 & 3787\\
		2 & 0. 3. 9. 11. 12. 13. 15. 17. 18. 19. 20. 22. 23. 25. 
		26. 27. 30. 32. 33. 35. 37. 41. 42. 44. 46. 47. 50. 51. 
		52. 55. 56. 58. 59. 60. 64. 69. 70. 72. 74. 75. 77. 80. 
		81. 85. 86. 88. 89. 90. 95. 96. 99. & 10365 & 3455\\
		3 & 0. 1. 2. 4. 6. 7. 11. 13. 14. 16. 19. 20. 21. 22. 24. 
		26. 28. 30. 31. 33. 34. 35. 37. 38. 39. 40. 41. 44. 45. 
		47. 48. 49. 50. 51. 53. 56. 61. 64. 68. 69. 71. 73. 76. 
		78. 81. 82. 86. 87. 89. 93. 94. 96. 97. 98. 99. & 13269 & 4424\\
		4 & 1. 2. 3. 4. 6. 8. 14. 17. 18. 21. 25. 27. 29. 31. 32. 33. 
		35. 40. 42. 46. 52. 54. 55. 56. 59. 60. 67. 68. 69. 70. 72. 74.
		75. 76. 78. 82. 83. 84. 85. 86. 88. 89. 91. 92. 93. 94. 95. 97. 
		98. 99. & 10005 & 3335\\
		\hline
	\end{tabular}
	\caption{Non-IID distribution on each client.}
	\label{Non-IID}    
\end{table*}

\begin{figure}[!t]
	\centering
	\includegraphics[width=0.45\textwidth]{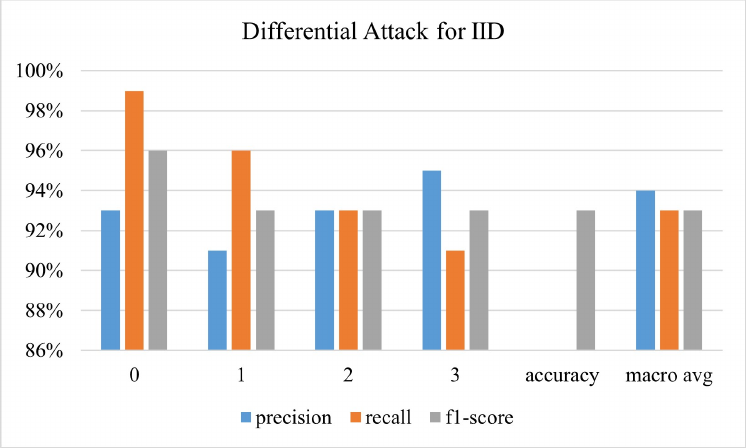}
	\caption{Differential Attack for IID.} \label{IID}
\end{figure}

\begin{figure}[!t]
	\centering
	\includegraphics[width=0.45\textwidth]{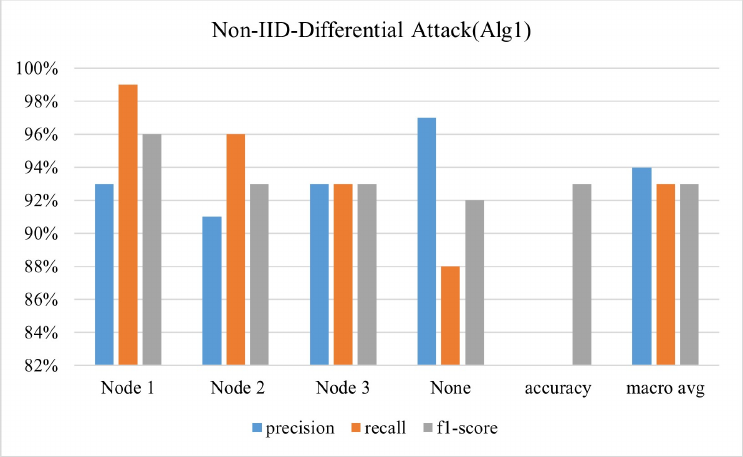}
	\caption{Differential Attack Algorithm 1 for Non-IID.} \label{NonIID1}
\end{figure}

Regarding the differential attack under independent identical distribution (IID), we achieve remarkably accurate results, as depicted in Figure~\ref{IID}. This experiment employs a distributed edge configuration with four clients. Figure~\ref{NonIID1} and Figure~\ref{NonIID2} present the results of differential attack algorithms 1 and 2, respectively, under the non-IID condition. These results are equally impressive. Even in our distributed edge system with four clients, we achieve 80\% accuracy, significantly surpassing the blind guess baseline (25\%). Notably, our distributed edge system exhibits lower resistance to differential attacks, particularly under non-IID conditions. Subsequently, the attack results under different aggregation weights are illustrated in Figure~\ref{NonIID2}.

\begin{figure}[!t]
	\centering
	\includegraphics[width=0.45\textwidth]{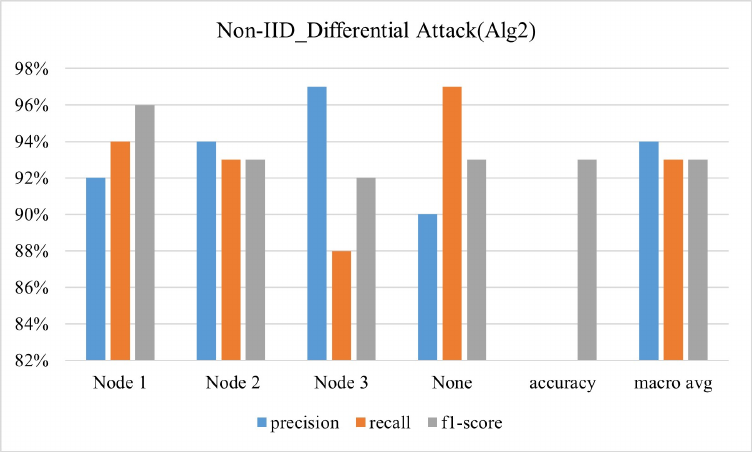}
	\caption{Differential Attack Algorithm 2 for Non-IID.} \label{NonIID2}
\end{figure}
\begin{figure}[!t]
	\centering
	\includegraphics[width=0.45\textwidth]{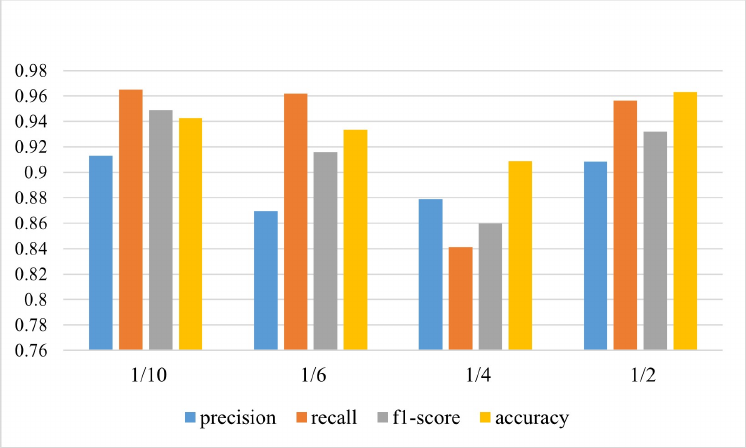}
	\caption{Differential Attack Algorithm 2 for Different Aggregation.} \label{NonIID2}
\end{figure}

\section{Defense Strategies}\label{Defense}
In this section, we employ several defense methods against membership inference attacks, including Regularization~\cite{DBLP:conf/ccs/NasrSH18},~\cite{DBLP:journals/corr/abs-2002-12062}, and Dropout~\cite{DBLP:conf/csfw/YeomGFJ18}, to assess their effectiveness in mitigating privacy risks. We utilize a distributed edge configuration with four clients and evaluate the defenses against the best-performing differential attack identified in previous experiments.

\begin{figure}[!t]
	\centering
	\includegraphics[width=0.45\textwidth]{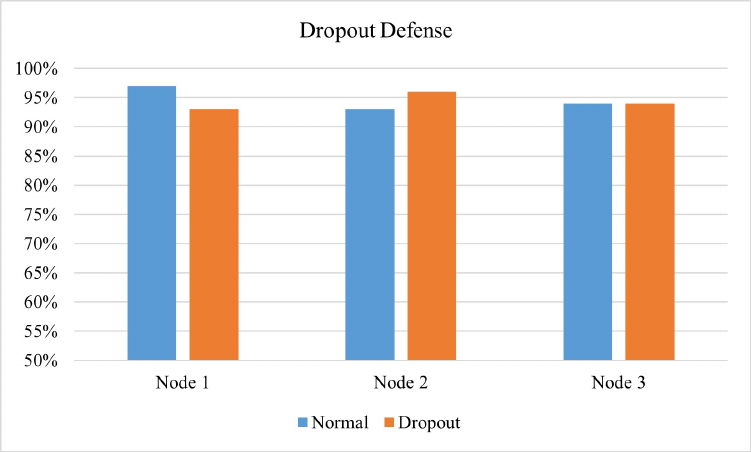}
	\caption{The results of the Dropout defense method.} \label{Dropout}
\end{figure}

\begin{figure}[!t]
	\centering
	\includegraphics[width=0.45\textwidth]{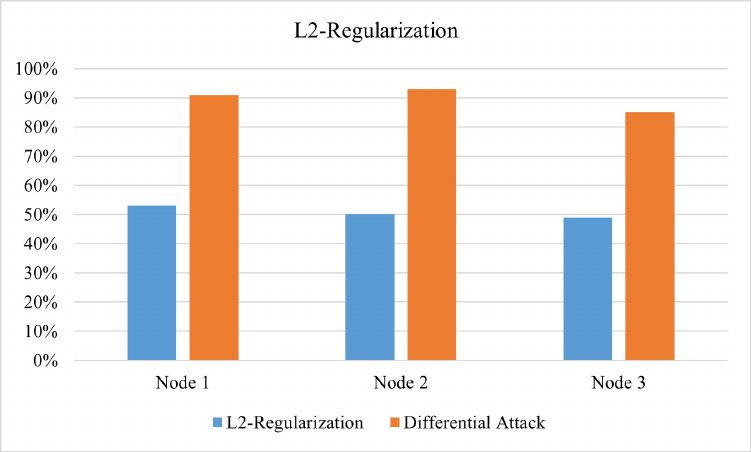}
	\caption{The results of the L2-Regularization defense method.} \label{Regularization}
\end{figure}

Specifically, we use the CIFAR-100 dataset with four clients in this section. For Dropout, we adopt the strategy outlined in~\cite{De2021}, incorporating dropout mechanisms after each max-pooling layer of the model. Each dropout mechanism is assigned a differential weight, typically 0.25 or 0.5. Regarding Regularization~\cite{DBLP:conf/ccs/NasrSH18}, we employ L2-Regularization with a weight of 0.001. Figure~\ref{Dropout} presents the attack results before and after the inclusion of dropout mechanisms, while Figure~\ref{Regularization} illustrates the defense effect of regularization. It's notable that regularization does not yield significant defense results. However, it appears that the differential attack itself may have a better defense against regularization, as evidenced by the improved defense results obtained when employing conventional attacks, as shown in Figure~\ref{Regularization}.


\section{Related Works}\label{Related}
\textbf{Distributed Edge Intelligence.}
Traditional machine learning methods rely heavily on massive data for training, posing challenges in data collection and privacy preservation during data sharing and transmission. Federated Learning (FL)\cite{DBLP:conf/sp/NasrSH19} offers a distributed learning approach to address this issue. However, FL still relies on a centralized server for model aggregation, leaving it vulnerable to information leakage during server-targeted attacks. In contrast, a novel distributed learning paradigm called swarm learning\cite{De2021,isa2023backdoor} empowers participating client users to drive model training. In swarm learning, there is no centralized server; instead, each user forms a decentralized network to transmit local model updates sequentially. Prior to each aggregation cycle, a client is randomly selected to act as a temporary server, enhancing system robustness by aggregating updates from all client users.

\textbf{Membership Inference Attack.} 
Following model training, machine learning models retain information about the training data, making them susceptible to a technique known as membership inference attack. Initially proposed for genomic data~\cite{pub.1013666372}, membership inference attacks have since demonstrated efficacy in contexts ranging from human mobility aggregation~\cite{DBLP:journals/corr/abs-1708-06145} to machine learning~\cite{DBLP:journals/corr/ShokriSS16}. Numerous recent studies have addressed this vulnerability~\cite{DBLP:journals/corr/abs-1806-01246,DBLP:conf/ccs/NasrSH18,DBLP:conf/sp/NasrSH19,DBLP:conf/icml/YurochkinAGGHK19,DBLP:journals/corr/abs-2101-01341,DBLP:conf/sp/NasrSH19,DBLP:conf/uss/SongM21,DBLP:conf/ccs/ChenYZF20}. For instance,\cite{DBLP:journals/corr/abs-1806-01246} proposed an enhanced NN-based membership inference attack and introduced metric attacks. Additionally,\cite{DBLP:conf/sp/NasrSH19} identified vulnerabilities in gradient descent and devised a white-box attack based on this insight. Moreover,~\cite{DBLP:journals/corr/abs-2101-01341} innovatively integrated the concept of differential comparison with membership inference attacks.

\section{Conclusion}\label{Conclusion}
In this study, we employ the membership inference attack to elucidate the latent data leakage inherent when edge nodes endeavor to collectively formulate a global model. Additionally, we address various defense mechanisms aimed at ameliorating this security vulnerability. It is our aspiration that this research will stimulate further investigations into data privacy within edge intelligence systems.


\end{document}